\documentclass[runningheads]{llncs}
\usepackage[T1]{fontenc}
\usepackage{graphicx}
\usepackage{multirow}
\usepackage{ulem}
\usepackage{threeparttable}
\usepackage{newtxmath}
\usepackage{makecell}
\usepackage{marvosym}
\usepackage[colorlinks=true, linkcolor=blue, citecolor=blue, urlcolor=blue]{hyperref}
\begin{document}
%
\title{ConvFormer: Plug-and-Play CNN-Style Transformers for Improving Medical Image Segmentation}
%
%
\author{Xian Lin\inst{1} 
	\and
	Zengqiang Yan\inst{1}\textsuperscript{(\Letter)} 
	\and
	Xianbo Deng\inst{2} 
	\and
	Chuansheng Zheng\inst{2} 
	\and
	Li Yu\inst{1} 
}
\authorrunning{X. Lin et al.}
%
\institute{School of Electronic Information and Communications, Huazhong University of Science and Technology \\
		\email{\{xianlin, z\_yan, hustlyu\}@hust.edu.cn}  \and
		Department of Radiology, Union Hospital, Tongji Medical College, Huazhong University of Science and Technology \\
		\email{dengxianbo@hotmail.com, cszheng@hust.edu.cn} 
	}
\maketitle              

\begin{abstract}
	
	Transformers have been extensively studied in medical image segmentation to build pairwise long-range dependence. Yet, relatively limited well-annotated medical image data makes transformers struggle to extract diverse global features, resulting in attention collapse where attention maps become similar or even identical. Comparatively, convolutional neural networks (CNNs) have better convergence properties on small-scale training data but suffer from limited receptive fields. Existing works are dedicated to exploring the combinations of CNN and transformers while ignoring attention collapse, leaving the potential of transformers under-explored. In this paper, we propose to build CNN-style Transformers (ConvFormer) to promote better attention convergence and thus better segmentation performance. Specifically, ConvFormer consists of pooling, CNN-style self-attention (CSA), and convolutional feed-forward network (CFFN) corresponding to tokenization, self-attention, and feed-forward network in vanilla vision transformers. In contrast to positional embedding and tokenization, ConvFormer adopts 2D convolution and max-pooling for both position information preservation and feature size reduction. In this way, CSA takes 2D feature maps as inputs and establishes long-range dependency by constructing self-attention matrices as convolution kernels with adaptive sizes. Following CSA, 2D convolution is utilized for feature refinement through CFFN. Experimental results on multiple datasets demonstrate the effectiveness of ConvFormer working as a plug-and-play module for consistent performance improvement of transformer-based frameworks. Code is available at \url{https://github.com/xianlin7/ConvFormer}.
	
\end{abstract}
\section{Introduction}
Benefiting from the prominent ability to model long-range dependency, transformers have become the de-facto standard for natural language processing~\cite{transformer}. Compared with convolutional neural networks (CNNs), which encourage locality, weight sharing, and translation equivariance, transformers build global dependency through self-attention layers, bringing more possibilities for feature exaction and breaking the performance ceiling of CNNs in return~\cite{unet,attunet,vit,setr,mae}.

Inspired by this, transformers are introduced into medical image segmentation and arouse wide concerns~\cite{caganformer,convfree,mmformer,smeswinunet,satr}. In vision transformers, each medical image is first split into a series of patches and then projected into a 1D sequence of patch embeddings~\cite{vit}. Through building pairwise interaction among patches/tokens, transformers are supposed to aggregate global information for robust feature exaction. However, learning well-convergence global dependency in transformers is highly data-intensive, making transformers less effective given relatively limited medical imaging data.

\begin{figure}[t]
	\centering
	\includegraphics[width=1\textwidth]{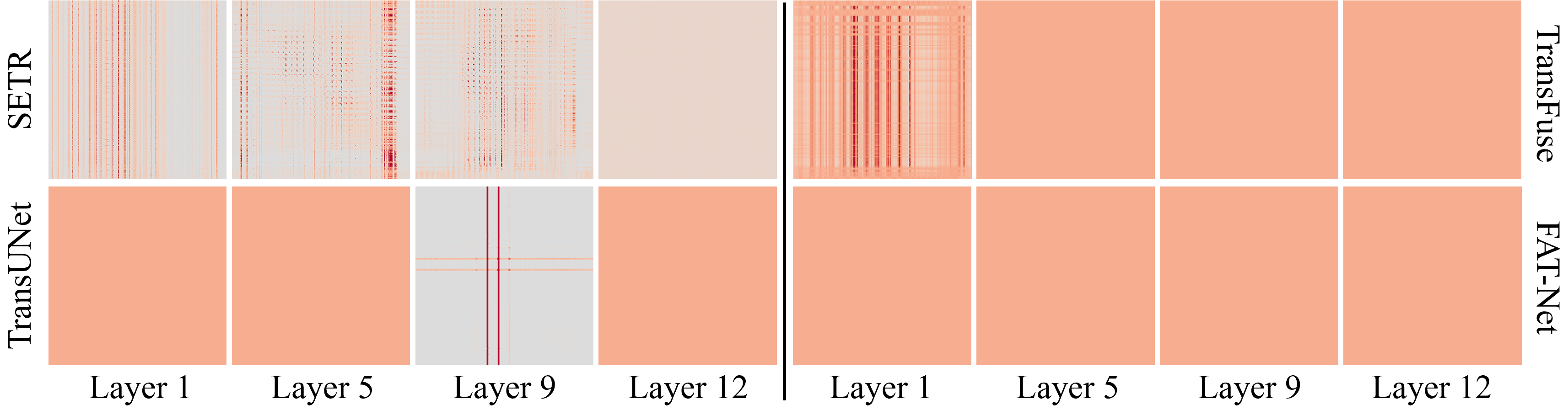}
	\caption{Visualization of attention maps from the selected layers of the first head in different transformer frameworks. The darker the color, the closer the dependency.}
	\label{fig1}
\end{figure}
To figure out how transformers work in medical image segmentation, we trained four state-of-the-art transformer-based models~\cite{setr,transunet,transfuse,fatnet} on the ACDC dataset and visualized the learned self-attention matrices across different layers as illustrated in Fig.~\ref{fig1}. For all approaches, the attention matrices tend to become uniform among patches (\textit{i.e.}, attention collapse~\cite{deepvit}), especially in deeper layers. Attention collapse is more noticeable, especially in CNN-Transformer hybrid approaches (\textit{i.e.}, TransUNet, TransFuse, and FAT-Net). On the one hand, insufficient training data would make transformers learn sub-optimal long-range dependency. On the other hand, directly combining CNNs with transformers would make the network biased to the learning of CNNs, as the convergence of CNNs is more achievable compared to transformers, especially on small-scale training data. Therefore, how to address attention collapse and improve the convergence of transformers is crucial for performance improvement.

In this work, we propose a plug-and-play module named ConvFormer to address attention collapse by constructing a kernel-scalable CNN-style transformer. In ConvFormer, 2D images can directly build sufficient long-range dependency without being split into 1D sequences. Specifically, corresponding to tokenization, self-attention, and feed-forward network in vanilla vision transformers, ConvFormer consists of pooling, CNN-style self-attention (CSA), and convolutional feed-forward network (CFFN) respectively. For an input image/feature map, its resolution is first reduced by applying convolution and max-pooling alternately. Then, CSA builds appropriate dependency for each pixel by adaptively generating a scalable convolutional, being smaller to include locality or being larger for long-range global interaction. Finally, CFFN refines the features of each pixel by applying continuous convolutions. Extensive experiments on three datasets across five state-of-the-art transformer-based methods validate the effectiveness of ConvFormer, outperforming existing solutions to attention collapse.

\section{Related Work}

Recent transformer-based approaches for medical image analysis mainly focus on introducing transformers for robust features exaction in the encoder, cross-scale feature interactive in skip connection, and multifarious feature fusion in the decoder~\cite{missformer,swinunet,levit,phtrans}. The study about addressing attention collapse for transformers in medical imaging is under-explored. Even in natural image processing, attention collapse, usually existing in the very deep layers of deep transformer-based models, has not been fully studied. Specifically, Zhou \textit{et al}. \cite{deepvit} developed Re-attention to re-generate self-attention matrices aiming at increasing their diversity on different layers. Zhou \textit{et al}. \cite{refiner} projected self-attention matrices into a high-dimensional space and applied convolutions to promote the locality and diversity of self-attention matrices. Touvron \textit{et al}. \cite{cait} proposed to re-weight the channels of the outputs from the self-attention module and the feed-forward module to facilitate the convergence of transformers.

\section{Method}

\begin{figure}[t]
	\centering
	\includegraphics[width=1\textwidth]{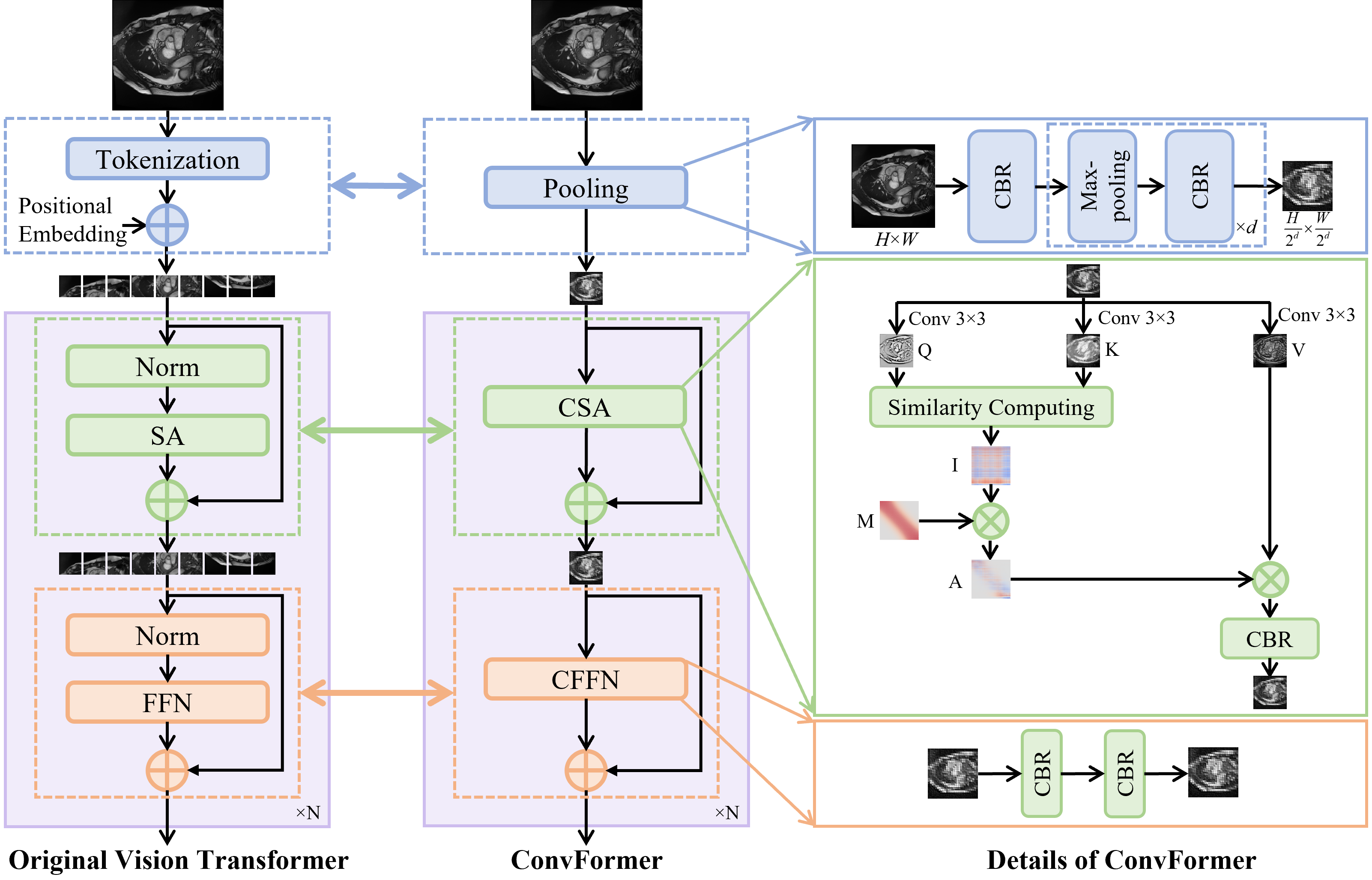}
	\caption{Comparison between vanilla vision transformer and ConvFormer. CBR is short for the combination of convolution, batch normalization, and Relu. Multiple heads are omitted for simplicity.}
	\label{fig2}
\end{figure}

The comparison between the vision transformer (ViT) and ConvFormer is illustrated in Fig.~\ref{fig2}. The greatest difference is that our ConvFormer is conducted on 2D inputs while ViT is applied to 1D sequences. Specifically, the pooling module is utilized to replace tokenization in ViT, which well preserves locality and positional information without extra positional embeddings. The CNN-style self-attention (CSA) module, \textit{i.e.} the core of ConvFormer, is developed to replace the self-attention (SA) module in ViT to build long-range dependency by constructing self-attention matrices in a similar way like convolutions with adaptive and scalable kernels. The convolutional feed-forward network (CFFN) is developed to refine the features for each pixel corresponding to the feed-forward network (FFN) in ViT. No upsampling procedure is adopted to resize the output of ConvFormer back to the input size as the pooling module can match the output size by adjusting the maxpooling times. It should be noticed that ConvFormer is realized based on convolutions, which eliminates the training tension between CNNs and transformers as analyzed in Section 1. Each module of ConvFormer is described in the following.

\subsection{Pooling vs. Tokenization}
The pooling module is developed to realize the functions of tokenization (\textit{i.e.}, making the input suitable to transformers in the channel dimension and shaping and reducing the input size when needed) while without losing details in the grid lines in tokenization. For an input $X_{in} \in \mathbb{R} ^ {c \times H \times W}$, convolution with a kernel size of $3\times3$ followed by batch normalization and Relu, is first applied to capture local features. Then, corresponding to each patch size $S$ in ViT, total $d = \log_2S$ downsampling operations are applied in the pooling module to produce the same resolutions. Here, each downsampling operation consists of a max-pooling with a kernel size of $2\times2$ and a combination of $3\times3$ convolution, batch normalization, and Relu. Finally, $X_{in}$ becomes $X_{1} \in \mathbb{R} ^ {c_m \times \frac{H}{2^d} \times \frac{W}{2^d}}$ through the pooling module where $c_m$ is corresponding to the embedding dimension in ViT.

\subsection{CNN-style vs. Sequenced Self-attention}

The building of long-range dependency in ConvFormer is relying on CNN-style self-attention, which creates an adaptive receptive field for each pixel by constructing a customized convolution kernel. Specifically, for each pixel $x_{i, j}$ of $X_1$, the convolution kernel $A^{i, j}$ is constructed based on two intermediate variables:
\begin{equation}
	Q_{i, j} = \sum_{l=-1}^{1}\sum_{g=-1}^{1}E^q_{2+l, 2+g}x_{i+l,j+g},
\end{equation}
\begin{equation}
	K_{i, j} = \sum_{l=-1}^{1}\sum_{g=-1}^{1}E^k_{2+l, 2+g}x_{i+l,j+g},
\end{equation}
where $E^q$ and $E^k \in \mathbb{R} ^ {c_q \times c_m\times 3 \times 3}$ are the learnable projection matrices and $c_q$ is corresponding to the embedding dimension of $Q$, $K$, and $V$ in ViT, which incorporates the features of adjacent pixels in $3\times3$ neighborhood into $x_{i, j}$. Then, the initial customized convolutional kernel $I^{i, j}\in \mathbb{R} ^ {\frac{H}{2^d} \times \frac{W}{2^d}}$ for $x_{i, j}$ is calculated by computing the cosine similarity:
\begin{equation}
	I^{i, j}_{m, n} = \frac{\sum_{l=0}^{c_q}Q_{i, j}K_{m, n}}{\sqrt{\sum_{l=0}^{c_q}Q_{i, j}^2}\sqrt{\sum_{l=0}^{c_q}K_{m, n}^2}}.
\end{equation}
Here, $I^{i, j}_{m, n} \in [-1, 1]$ and seldom occurs $I^{i, j}_{m, n} = 0$. $I^{i, j}_{m,n}$ corresponds to attention score calculation in ViT (constrained to be positive while $I^{i, j}_{m, n}$ can be either positive or negative). Then, we dynamically determine the size of the customized convolution kernel for $x_{i, j}$ by introducing a learnable Gaussian distance map $M$:
\begin{equation}
	M^{i, j}_{m, n} = e^{-\frac{(i -m)^2(2^d/H)^2+(j-n)^2(2^d/W)^2}{2(\theta \times \alpha)^2}},
\end{equation}
where $\theta \in (0, 1)$ is a learnable network parameter to control the receptive field of $A$ and $\alpha$ is a hyper-parameter to control the tendency of the receptive field. $\theta$ is proportional to the receptive field. For instance, under the typical setting $H=W=256$, $d=3$, and $\alpha=1$, when $\theta=0.003$, the receptive field only covers five adjacent pixels, when $\theta>0.2$, the receptive field is global. The larger $\alpha$ is, the more likely $A$ tends to have a global receptive field. 
Based on $I^{i, j}$ and $M^{i, j}$, $A^{i, j}$ is calculated by $A^{i, j}=I^{i, j} \times M^{i, j}$. In this way, every pixel $x^{i, j}$ has a customized size-scalable convolution kernel $A^{i, j}$. By multiplying $A$ with $V$, CSA can build adaptive long-range dependency, where $V$ can be formulated similarly according to Eq. (1). Finally, the combination of $1\times1$ convolution, batch normalization, and Relu is utilized to integrate features learned from long-range dependency.

\subsection{Convolution vs. Vanilla Feed-forward Network}

The convolution feed-forward network (CFFN) is to refine the features produced by CSA, just consisting of two combinations of $1\times1$ convolution, batch normalization, and Relu. By replacing linear projection and layer normalization in ViT, CFFN makes ConvFormer completely CNN-based, avoiding the combat between CNN and Transformer during training like CNN-Transformer hybrid approaches.

\section{Experiments}
\subsection{Datasets and Implementation Details}

\noindent \textbf{ACDC}$\footnote[1]{https://www.creatis.insa-lyon.fr/Challenge/acdc/}$. A publicly-available dataset for the automated cardiac diagnosis challenge. Totally 100 scans with pixel-wise annotations of left ventricle (LV), myocardium (MYO), and right ventricle (RV) are available~\cite{acdc}. Following \cite{transunet,swinunet,levit}, 70, 10, and 20 cases are used for training, validation, and testing respectively.

\noindent  \textbf{ISIC 2018}$\footnote[2]{https://challenge.isic-archive.com/data/}$. A publicly-available dataset for skin lesion segmentation. Totally 2594 dermoscopic lesion images with pixel-level annotations are available~\cite{isic1,isic2}. Following \cite{dstransunet,transattunet}, the dataset is randomly divided into 2076 images for training and 520 images for testing.

\noindent  \textbf{ICH.} A locally-collected dataset for hematoma segmentation. Totally 99 CT scans consisting of 2648 slices were collected and annotated by three radiologists. The dataset is randomly divided into the training, validation, and testing sets according to a ratio of 7:1:2.
\begin{table}[!t]
	\centering
	\caption{Quantitative results in Dice (DSC)) and Hausdorff Distance (HD).}\label{tab1}
	\begin{threeparttable}
		\begin{tabular}{l|cccc|c|c|c|c|c}
			\hline
			\multirow{3}{*}{Method} & \multicolumn{6}{c|}{DSC (\%) $\uparrow$}  &  \multicolumn{3}{c}{HD (\%) $\downarrow$}    \\ \cline{2-10}
			& \multicolumn{4}{c|}{ACDC}       & \multirow{2}{*}{ISIC}  & \multirow{2}{*}{ICH} & ACDC    & \multirow{2}{*}{ISIC}  & \multirow{2}{*}{ICH}   \\ \cline{2-5} \cline{8-8}
			& Avg.                     & RV             & MYO                & LV                         &               &                    & Avg.                       &                &            \\ \hline
			SETR~\cite{setr}                    & 87.14                              & 83.95                              & 83.89                              & 93.59                      & 89.03          & 80.17          & 17.24     & 22.33    & 14.90\\
			+Re-attention~\cite{deepvit}       & 85.91                              & 82.52                              & 82.27                              & 92.93                      & 88.00          & 79.08          & 18.21    & 24.57     & 14.50\\
			+LayerScale~\cite{cait}         & 85.74                              & 81.54                              & 82.70                              & 92.98                      & 87.98          & 78.91          & 17.88    & 23.94     & 14.45\\
			+Refiner~\cite{refiner}            & 85.75                              & 83.18                              & 81.61                              & 92.46                      & 86.63          & 78.35          & 18.09    & 25.43     & 14.95\\
			+ConvFormer         & \textbf{91.00}                     & \textbf{89.26}                     & \textbf{88.60}                     & \textbf{95.15}             & \textbf{90.41$^\star$} & \textbf{81.56} & \textbf{14.08}    & \textbf{21.68}     & \textbf{13.50}\\ \hline
			TransUNet~\cite{transunet}               & 90.80                              & 89.59                              & 87.81                              & 94.99                      & 88.75          & 78.52          & 14.58    & 25.11     & 15.90\\
			+Re-attention~\cite{deepvit}  & 91.25                              & 89.91                              & 88.61                              & 95.22                      & 88.35          & 77.50          & 13.79    & \textbf{23.15}     & \textbf{15.40}\\
			+LayerScale~\cite{cait}    & 91.30                              & 89.37                              & 88.79                              & \textbf{95.75}             & 88.75          & 75.60          & \textbf{13.68}    & 23.32     & 15.50\\
			+Refiner~\cite{refiner}       & 90.76                              & 88.66                              & 88.39                              & 95.22                      & 87.90          & 76.47          & 14.73    & 25.31     & 15.65\\
			+ConvFormer    & \textbf{91.42} & \textbf{90.17} & \textbf{88.84} & 95.25 & \textbf{89.40} & \textbf{80.66} & 13.96 & 23.19 & 15.70\\ \hline
			TransFuse~\cite{transfuse}               & 89.10          & 87.85          & 85.73          & 93.73 & 89.28          & 75.11          & 14.98 & 23.08 & 18.60\\
			+Re-attention~\cite{deepvit}  & 88.48          & 87.05          & 85.37          & 93.00 & 88.28          & 73.74    & 16.20        & 24.56 & 18.45     \\
			+LayerScale~\cite{cait}    & 88.85         & 87.81          & 85.24          & 93.50 & 89.00          & 74.18        & 13.53 & 23.96 & 20.00 \\
			+Refiner~\cite{refiner}       & 89.06          & 87.88          & 85.55          & 93.75 & 85.65          & 74.16    & 14.05 & 26.30 & 18.60     \\
			+ConvFormer    & \textbf{89.88}                     & \textbf{88.85}                     & \textbf{86.50}                     & \textbf{94.30}             & \textbf{90.56$^\star$} & \textbf{75.56} & \textbf{12.84} & \textbf{21.30} & \textbf{17.60}\\ \hline
			FAT-Net~\cite{fatnet}                 & 91.46                              & 90.13                              & 88.61                              & 95.60                      & 89.72          & 83.73          & 13.82 & 22.63 & 16.20 \\
			+Re-attention~\cite{deepvit}    & 91.61$^\star$                              & 89.99                              & 89.18                              & 95.64                      & 89.84          & 84.42          & 14.00 & 22.54 & 14.20 \\
			+LayerScale~\cite{cait}      & 91.71$^\star$                              & 90.01                              & 89.39                              & 95.71                      & 90.06          & 83.87          & 13.50 & 21.93 & 13.70 \\
			+Refiner~\cite{refiner}         & 91.94$^\star$                              & 90.54                              & \textbf{89.70}                     & 95.58                      & 89.20          & 83.14          & 13.37 & 23.35 & \textbf{13.25} \\
			+ConvFormer      & \textbf{92.18$^\star$}                     & \textbf{90.69}                     & 89.57                              & \textbf{96.28}             & \textbf{90.36$^\star$} & \textbf{84.97} & \textbf{11.32} & \textbf{21.73} & 14.10\\ \hline
			Patcher~\cite{patcher}                 & 91.41                              & 89.56                              & 89.12                              & 95.53                      & 89.11          & 80.54          & 13.55 & 22.16 & 15.70\\
			+Re-attention~\cite{deepvit}    & 91.25                              & 89.77                              & 88.58                              & 95.39                      & 89.73          & 79.08          & 14.49 & 21.86 & 17.60\\
			+LayerScale~\cite{cait}      & 91.07                              & 88.94                              & 88.81                              & 95.46                      & 90.16          & 74.13          & 15.48 & \textbf{21.78} & 18.60\\
			+Refiner~\cite{refiner}         & 91.26                              & 89.65                              & 88.58                              & 95.57                      & 68.92          & 79.66          & 13.93 & 56.62 & 15.60\\
			+ConvFormer       & \textbf{92.07$^\star$}                     & \textbf{90.91}                     & \textbf{89.54}                     & \textbf{95.78}             & \textbf{90.18} & \textbf{81.69} & \textbf{12.29} & 21.88 & \textbf{15.35}\\ \hline
		\end{tabular}
		\begin{tablenotes}
			\item[$\star$] Approaches outperforming the state-of-the-art 2D approaches on the publicly-available ACDC (\textit{i.e.}, FAT-Net~\cite{fatnet}: 91.46\% in Avg. DSC) and ISIC (\textit{i.e.}, Ms Red~\cite{msred}: 90.25\% in Avg. DSC) datasets respectively. \textit{More comprehensive quantitative comparison results can be found in the supplemental materials.}
		\end{tablenotes}
	\end{threeparttable}
\end{table}

\noindent \textbf{Implementation Details.} For a fair comparison, all the selected state-of-the-art transformer-based baselines were trained with or without ConvFormer under the same settings. All models were trained by an Adam optimizer with a learning rate of 0.0001 and a batch size of 4 for 400 rounds. Data augmentation includes random rotation, scaling, contrast augmentation, and gamma augmentation.

\subsection{Results}

ConvFormer can work as a plug-and-play module and replace the vanilla transformer blocks in transformer-based baselines. To evaluate the effectiveness of ConvFormer, five state-of-the-art transformer-based approaches are selected as backbones, including SETR~\cite{setr}, TransUNet~\cite{transunet}, TransFuse~\cite{transfuse}, FAT-Net~\cite{fatnet}, and Patcher~\cite{patcher}. SETR and Patcher utilize pure-transformer encoders, while TransUNet, TransFuse, and FAT-Net adopt CNN-Transformer hybrid encoders. In addition, three state-of-the-art methods for addressing attention collapse, including Re-attention~\cite{deepvit}, LayerScale~\cite{cait}, and Refiner~\cite{refiner}, are equipped with the above transformer-based baselines for comparison.

\noindent \textbf{Quantitative Results.} Quantitative results of ConvFormer embedded into various transformer-based baselines on the three datasets are summarized in Table~\ref{tab1}. ConvFormer achieves consistent performance improvements on all five backbones. Compared to CNN-Transformer hybrid approaches (\textit{i.e.}, TransUNet, TransFuse, and FAT-Net), ConvFormer is more beneficial on pure-transformer approaches (\textit{i.e.}, SETR and Patcher). Specifically, with ConvFormer, SETR achieves an average increase of 3.86$\%$, 1.38$\%$, and 1.39$\%$ in Dice on the ACDC, ISIC, and ICH datasets respectively, while the corresponding performance improvements of Patcher are 0.66$\%$, 1.07$\%$, and 1.15$\%$ respectively. Comparatively, in CNN-Transformer hybrid approaches, as analyzed above, CNNs would be more dominating against transformers during training. Despite this, re-balancing CNNs and Transformers through ConvFormer can build better long-range dependency for consistent performance improvement.

\noindent \textbf{Comparison with SOTA Approaches.} Quantitative results compared with the state-of-the-art approaches to addressing attention collapse are summarized in Table~\ref{tab1}. In general, given relatively limited training data, existing approaches designed for natural image processing are unsuitable for medical image segmentation, resulting in unstable performance across different backbones and datasets. Comparatively, ConvFormer consistently outperforms these approaches and brings stable performance improvements to various backbones across datasets, demonstrating the excellent generalizability of ConvFormer as a plug-and-play module.

\begin{figure}[t]
	\centering
	\includegraphics[width=\textwidth]{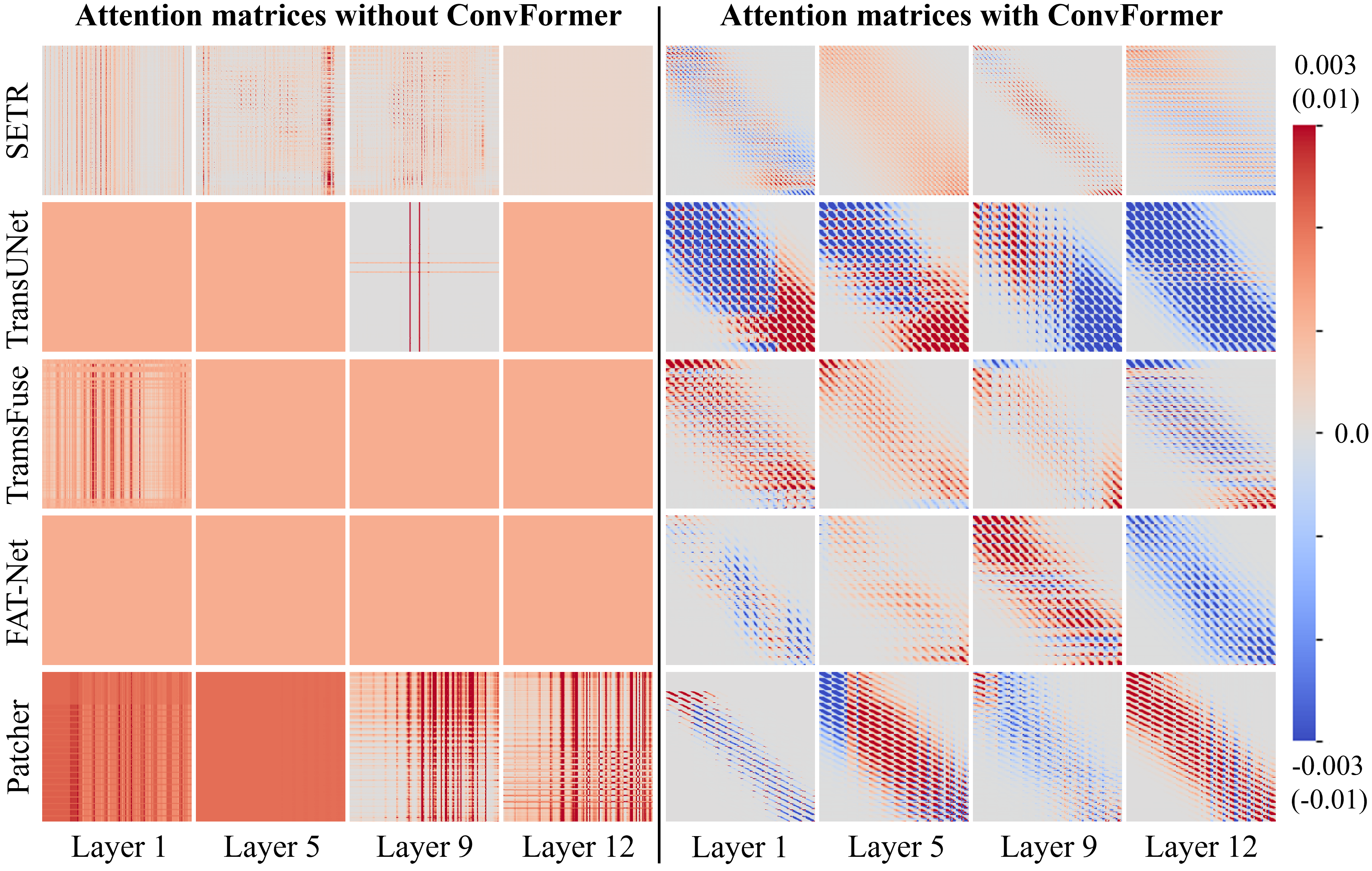}
	\caption{Visualization of self-attention matrices by baselines w/ and w/o ConvFormer.}
	\label{fig3}
\end{figure}

\noindent  \textbf{Visualization of Self-Attention Matrices.} To qualitatively evaluate the effectiveness of ConvFormer in addressing attention collapse and building efficient long-range dependency, we visualize the self-attention matrices with and without ConvFormer as illustrated in Fig.~\ref{fig3}. By introducing ConvFormer, attention collapse is effectively alleviated. Compare to the self-attention matrices of baselines, the matrices learned by ConvFormer are more diverse. Specifically, the interactive range for each pixel is scalable, being small for locality preserving or being large for global receptive fields. Besides, dependency is no longer constrained to be positive like ViT, which is more consistent with convolution kernels. \textit{Qualitative segmentation results of different approaches on the three datasets can be found in the supplemental materials.}

\noindent \textbf{Ablation Study} As described in Sec. 3.2, $\alpha$ is to control the receptive field tendency in ConvFormer, The larger the $\alpha$, the more likely ConvFormer contains larger receptive fields. To validate this, we conduct an ablation study on $\alpha$ as summarized in Table~\ref{tab2}. In general, using a large $\alpha$ does not necessarily lead to more performance improvements, which is consistent with our observation that not every pixel needs global information for segmentation.

\begin{table}[!t]
	\centering
	\caption{Ablation study of hyper-parameter $\alpha$ on the ACDC dataset.}\label{tab2}
	\begin{tabular}{p{1.6cm}<{\centering}|p{1.6cm}<{\centering}p{1.6cm}<{\centering}p{1.6cm}<{\centering}p{1.6cm}<{\centering}p{1.6cm}<{\centering}}
		\hline
		$\alpha$ & 0.2   & 0.4   & 0.6   & 0.8 & 1.0   \\ \hline
		Dice ($\%$)  & 90.71 & \textbf{91.00} & 90.76 &  90.66   & 90.45 \\ \hline
	\end{tabular}
\end{table}

\section{Conclusions}

In this paper, we construct the transformer as a kernel-scalable convolution to address the attention collapse and build diverse long-range dependencies for efficient medical image segmentation. Specifically, it consists of pooling, CNN-style self-attention (CSA), and convolution feed-forward network (CFFN). The pooling module is first applied to extract the locality details while reducing the computational costs of the following CSA module by downsampling the inputs. Then, CSA is developed to build adaptive long-range dependency by constructing CSA as a kernel-scalable convolution, Finally, CFFN is used to refine the features of each pixel. Experimental results on five state-of-the-art baselines across three datasets demonstrate the prominent performance of ConvFormer, stably exceeding the baselines and comparison methods across three datasets.
\\
\\
\textbf{Acknowledgement.} This work was supported in part by the National Natural Science Foundation of China under Grant 62271220 and Grant 62202179, and in part by the Natural Science Foundation of Hubei Province of China under Grant 2022CFB585. The computation is supported by the HPC Platform of HUST.

\end{document}